\documentstyle[preprint,aps]{revtex}
\begin{document}
\title{ Quarkonium spectra with a Klein-Gordon Oscillator as a
confining potential}
\author{A. G. Grunfeld \thanks{Electronic address: 
grunfeld@venus.fisica.unlp.edu.ar}
and M. C. Rocca\\
{\it Departamento de F\'{\i}sica, Fac. de Cs. Exactas,
Universidad Nacional de La Plata.\\
C.C. 67 (1900) La Plata. Argentina.}}
\maketitle
\thispagestyle{empty}

\begin{abstract}
A quantum relativistic model is proposed to
represent the confining potential of $c \bar{c}$ and $b \bar{b}$ mesons
by using the Klein Gordon oscillator. We have solved the two-body
problem obtaining the eigenvalues and the
corresponding mass spectra, which fit very well with experimental data.
We compare the results of our model with others obtained from
different quark binding potentials.
\end{abstract}
\vspace{1cm}

\newpage

\section{INTRODUCTION}
The study of bound sates of quark antiquark systems has 
been performed with different formalisms in the last two decades. 
These systems are a profitable testing ground for new ideas 
and methods to explain confinement and strong interaction dynamics.
In particular, the mass spectrum for charmed mesons has largely
been measured and modeled. 
Some time ago, Kang and Schnitzer \cite{Kang} have proposed a model 
to simulate confinement between the constituents of mesons.
These authors considered a linear potential (LP) in the Klein Gordon 
equation. In this way they obtained the mass spectrum for charmonium.
Later on, Ram and Hasala \cite{Ram} followed this procedure with a 
quadratic potential (QP). 
There are also some works in the literature, devoted to the study of
quarkonium systems from a non-relativistic point of view, with different
potentials. As an example of that Eichten et al. \cite{Eich} have 
studied the charmonium family with linear plus Coulomb interaction
as a confining potential. Furthermore, Quigg and Rosner \cite{Quigg}
have studied the bound states of quarkonium with logarithmic, and power
law potentials. \\
\indent In this work we introduce an alternative relativistic quantum 
mechanical model which includes the Klein Gordon Oscillator in order to 
obtain the mass spectra for bound states of $b \bar{b}$ and 
$c \bar{c}$ mesons. We present a scheme that considers quarks as 
spinless particles, in the same spirit of ref.\cite{Kang} and \cite{Ram},
which reproduces with high precision the charmonium 
and bottonium mass spectra. 
The harmonic oscillator (HO) is a natural candidate to simulate the quark 
binding. It is one of the simplest potentials, well
behaved, which takes into account asymptotic freedom and confinement.
The Klein-Gordon oscillator (KGO) was introduced by Bruce and 
Minning \cite{Bruce}. This interaction gives the usual HO in the 
non-relativistic limit. \\
\indent In this frame we have solved the two-body problem. 
This alternative relativistic treatment is very suitable to study
bound states due to its simplicity. 
We compare our results for charmonium, 
with those obtained for linear and quadratic potentials in the
Klein Gordon equation \cite{Kang} \cite{Ram}.
Moreover, we have included in our analysis the values for the mass
spectra for $c \bar{c}$ and $b \bar{b}$ mesons that 
we have obtained by considering the treatment 
for two body systems in a quantum relativistic frame. 
This treatment was presented in a paper by Arshansky and Horwitz \cite{Hor}. 
There, it is derived the mass
spectra and the wave functions for the 4-D harmonic oscillator (4D) as a 
quantum relativistic two-body bound state, but the authors have not
fit their mass spectrum with the systems we are interested on, something 
we shall do below.\\
\indent The paper is organized as follows. In Section II we comment on the
Klein-Gordon oscillator formalism. Section III is devoted to describe
the model we propose. In Sect. IV we compare our results for charmonium
with those obtained in refs. \cite{Kang}\cite{Ram}, and with those that  
we have obtained considering the formalism in ref.\cite{Hor}. We also
present a table for the results of bottonium masses within 
KGO and 4D models. Finally, we sketch our conclusions.

\section{The formalism}

In 1989 Moshinsky and Szczepaniak \cite{Mosh} proposed a new type 
of interaction in the Dirac equation. This potential is linear in
both, momenta and coordinates, and the corresponding equation was named 
``Dirac Oscillator'' because in the non-relativistic limit, the harmonic
oscillator is obtained. Later on, this kind of interaction was also 
introduced in the Klein-Gordon (K-G) equation \cite{Bruce} by making
the minimal substitution:
\begin{equation}
\vec{P}  \rightarrow \vec{P} - i m \hat{\gamma} \hat{\Omega} . \vec{Q} 
\label{000}
\end{equation}
where $m$ is the mass of the particle
and
$\vec{P} = \hat{\eta}\vec{p}$ ,
$\vec{Q} = \hat{\eta}\vec{q}$  
are respectively, coordinate and momentum operators. 
The matrix $\hat{\Omega}$ is chosen to be 3x3 with coefficients
$ \hat{\Omega}_{ij} = {\omega}_i {\delta}_{ij} $. The constants ${\omega}_i$
(with i = 1, 2, 3) are the oscillator frequencies along the $x, y, z$ axes.
The matrix $\hat{\gamma} $
satisfies the anticommutation relations:
\[  \{ \hat{\gamma} , \hat{\eta} \} = 0  \;\;\;\;\;\;{\hat{\gamma}}^2 
= {\hat{\eta}}^2 = 1 .\]
Thus, the corresponding K-G equation (in natural units)
for the wave function $\Psi (\vec{q}, t)$ has the following form :
\begin{center}
\[ {\frac {{\partial}^2}{{\partial t}^2}} \Psi (\vec{q}, t) = 
\left( {\vec p}^2 + m^2 {\vec q} . {\vec{\Omega}}^2 . {\vec q} +
m {\hat{\gamma}} tr {\Omega} + m^2 \right) \Psi (\vec{q}, t). \]
\end{center}
In order to solve it, it is convenient to recast the K-G equation,
and some ways to do this are shown in ref.\cite{Val}.
One of these posibilities is the Sakata-Taketani approach \cite{Saka} 
characterized by the equation:
\begin{equation}
i{\frac {\partial}{\partial t}} \Phi =
\left\{ \frac{{\vec{p}} ({\tau}_3 + 
i{\tau}_2) {\vec{p}}}{2m} + m {\tau}_3 \right\} \Phi
\label{0011}
\end{equation}
where $ {\tau}_i $ are Pauli matrices and $\Phi$ has the form 
\begin{equation}
\Phi =
\left(
\begin{array}{rc}
\phi  \\
\chi
\end{array}
\right).
\end{equation}
This two component wave function is related with $\Psi$ as follows
\begin{equation}
\left\{
\begin{array}{rc}
\phi = \left( \Psi + \frac{i}{m} {\partial}_t \Psi \right) /2  \\
\chi = \left( \Psi - \frac{i}{m} {\partial}_t  \Psi \right) /2 .
\end{array}
\right.
\end{equation}
In order to obtain the oscillator-like
equation, in the isotropic case where $\omega_1 = \omega_2 = \omega_3$, 
the minimal coupling 
\begin{equation}
\vec{p} \rightarrow \vec{p} - im{\omega}{{\tau}_1}{\vec{r}}
\label{00i}
\end{equation}
must be introduced in the K-G equation (\ref{0011}) giving:
\begin{equation}
i{\frac {\partial}{\partial t}} \Phi =
\left\{ \frac{({\vec{p} - im{\omega}{\tau}_1}{\vec{r}}) 
({\tau}_3 + i{\tau}_2)({\vec{p} - im{\omega}{{\tau}_1}{\vec{r}})}}{2m} 
+ m {\tau}_3 \right\} \Phi
\label{001}
\end{equation}
which yields 
\begin{equation}
\left[ \Box + 2 \, m \, \omega^2 \, r^2 \, - \, 3 \, m \, \omega \, 
+ m^2 \right]  \Psi \; = \;0.  
\label{00122}
\end{equation}
The corresponding energy spectrum \cite{Val} is
\begin{equation}
E^2 = 2 m \omega \left( N_1 + N_2 + N_3 \right) + m^2
\end{equation}
with $N_1, N_2, N_3 = 0, 1...$\\
\indent On the basis of the results presented in this summary,
our model can be introduced.
Our purpose is to work out the two-body problem 
in the frame of the formalism we have commented.

\section{THE MODEL}

We propose that the above KG oscillator could model a quark antiquark
confining potential taking into account some further considerations.
In the case of one body KGO, we note from (\ref{00i}) that the oscillator
interaction has its origin at 0. If one desires to change the origin 
to $r_0$, the  replacement $r - r_0$ must be done.
We are interested in studying the case of two interacting particles with the
same mass. We call with subindex 1 and 2 the
magnitudes corresponding to each particle.
We propose a slight different ansatz when compared to (\ref{00i}). It reads,
\begin{equation}
 \vec{p_1} \rightarrow 
{\vec{p_1}}' = \vec{p_1} - im {\frac{{\omega}}{2}}{{\tau}_1}
({\vec{r_1}} - {\vec{r_2}}) 
\end{equation}
\begin{equation}
\vec{p_2} \rightarrow 
{\vec{p_2}}' = \vec{p_2} 
- im {\frac{{\omega}}{2}}{{\tau}_1}({\vec{r_2}} - {\vec{r_1}}),
\label{002i}
\end{equation}
giving rise to two coupled equations:
\begin{equation}
i\frac {\partial}{\partial t_1} \Phi =
\left[ \frac{\vec{p_1}' (\tau_3 + i\tau_2) \vec{p_1}'}{2m} 
+ m \tau_3 \right] \Phi
\label{0013a}
\end{equation}
and
\begin{equation}
i\frac {\partial}{\partial t_2} \Phi =
\left[ \frac{\vec{p_2}' 
(\tau_3 + i\tau_2) \vec{p_2}'}{2m} + m \tau_3 \right] \Phi.
\label{0013b}
\end{equation}

Summing the last two expressions, one obtains
an equation which corresponds to the case of two particles:

\begin{equation}
i \frac {\partial}{\partial t_1} \Phi + i \frac {\partial}{\partial t_2}\Phi =
\left[ \frac{\vec{p_1}' (\tau_3 + i\tau_2) \vec{p_1}'}{2m} +
\frac{\vec{p_2}' (\tau_3 + i\tau_2) \vec{p_2}'}{2m}
+ 2 m \tau_3 \right] \Phi.
\label{0013}
\end{equation}
In quantum mechanics, the study of a two body system could be thought of as
a free particle plus a particle in a central potential,
which depends on the relative coordinate $ \vec{r} = \vec{r_1} - \vec{r_2}$.
In order to solve our system we have made the following change of
variables:
\begin{equation}
T = \frac{t_1 + t_2}{2}, \hspace{3cm} t = t_1 - t_2 
\end{equation}
and 
\begin{equation}
\vec{R} = \frac{\vec{r}_1 + \vec{r}_2}{2}, 
\hspace{3cm} \vec{r} = \vec{r}_1 - \vec{r}_2. 
\end{equation}
The momenta $p_1$ and $p_2$ can be written as follows:
\begin{equation}
{\vec{p}}_1 = \frac{\vec{P}}{2} + {\vec{p}}
\end{equation}
\begin{equation}
{\vec{p}}_2 = \frac{\vec{P}}{2} - {\vec{p}}.
\end{equation}
Shifting now to these new variables, eq.(\ref{0013}) gives:
\begin{equation}
\left( \frac{{\partial}^2}{{\partial T}^2} + P^2 + 4 p^2 + 
m^2 \omega^2 r^2 - 6 m \omega + 4 m^2 \right) 
\Phi({\vec{P}}, {\vec{p}}, T).
\end{equation}
This equation can be solved by separation: 
\begin{equation}
\Phi({\vec{R}}, {\vec{r}}, T) =  
{\phi}_1(\vec{R}, T). {\phi}_2(\vec{r})
\end{equation}
giving rise to the following two equations 
\begin{equation}
\left( \frac{{\partial}^2}{{\partial T}^2} + P^2 + E^2 \right) 
{\phi}_1(\vec{R}, T) = (\Box + E^2) {\phi}_1(\vec{R}, T)
\end{equation}
and
\begin{equation}
\left({\vec{p}}^2 +  \mu^2 {\omega}^2 r^2 - 3 \mu \omega + 4\mu^2
- \frac{E^2}{4} \right) {\phi}_2(\vec{r}) = 0
\end{equation}
where $m = 2\mu$.
We are particularly interested in solving the last equation,
since the first one, has of trivial solution.
In spherical coordinates this equation takes the form:
\begin{equation}
\left( {p_r}^2 + \frac{L^2}{r^2} + \mu^2 {\omega}^2 r^2 
- 3 \mu \omega + 4 \mu^2  - \frac{E^2}{4} \right) {\phi}_2(\vec{r}) = 0.
\end{equation}
The first term can be written as:
\begin{equation}
p_r = - i \left( \frac{\partial}{\partial r} + \frac{1}{r} \right).
\end{equation}
This Hermitian operator is the radial component of the linear momentum.
There are two independent eigenfunctions. In analogy with non-relativistic
quantum harmonic oscillator, only one of them is physically acceptable
because the other diverges at the origin. So, the eigenvalues and
eigenfunctions are:
\begin{equation}
E^2 - 16 \mu^2 = 8 \mu \omega ( 2n + l ) \;\;\;\;\;\; 
n = 0, 1... \;\;\;\;\; l = 0, 1...
\label{e}
\end{equation}
\begin{equation}
\Phi(r, \theta, \phi)_{n, l, m} = A_{n,l} r^l e^{\frac{-\mu \omega}{2} r^2}
{L^{l+ \frac{1}{2}}}_{n}(\mu \omega r^2)
{Y^l}_m(\theta, \phi) 
\end{equation}
and  $ A_{n,l} $ is the normalization constant:
\begin{equation}
A_{n,l} = \left[ \frac{2 (\mu \omega)^{l+ \frac{3}{2}} \Gamma (n+1)}
{\gamma (n+l+ \frac{3}{2})} \right]^{1/2}.
\end{equation}
Note that the states are degenerated for particular values of $n$ and $l$.
The same situation appears in the 4D model. \\
\indent The solution we have previously derived enable us to compute the mass
spectrum for quarkonium, as we will show below.

\section{RESULTS AND CONCLUSIONS}

Let us now compute some specific values for the mass spectra for 
charmonium and bottonium. We have considered a two free parameter set: 
$m_{q}$ and $\omega$. For the $c \bar{c}$ family $m_c$ = 1.5 GeV
and $\omega \simeq$ 0.385 GeV.
Instead, for bottonium, $m_b$ = 5.17 Gev and $\omega \simeq$ 0.12 GeV.
The values for $\omega$ were obtained by adjusting
the rms radius of the ground state \cite{Eich,Blai}. \\
\indent Our results are shown in Tables I and II. Table I displays the values
for the masses of charmonium mesons within different relativistic
models, for particular quantum numbers $n$ and $l$. The explicit
values for the 4D model arise from the computation of charmonium
spectrum in such a general scheme. The other two known values, LP and QP,
are taken from ref.\cite{Kang,Ram}. With bold typos (KGO) are 
presented the values we have obtained with our model. 
From Table I, we can see that our scheme produces acceptable
results, without any further assumption. \\
\indent In Table II, results for $b \bar{b}$ mesons are presented. Here, 
only the 4D model can be confronted with our one, because, as far as we know,
the LP and QP treatments were not applied to this mesons. To do such a
comparison we choose the mass for bottom quark to be 5.16 GeV, which
is bigger than the experimental value. This value for the mass was
used in \cite{Eich} yielding $\omega \simeq$ 0.113 GeV.
Thus, we can expect an overestimation of the resulting mass spectrum,
due to such a choice.  
Anyway, taken the $b$ mass heavier than the experimental value, 
the masses for $b\bar{b}$ mesons predicted from these two models are
very close to the experimental data.
We again get agreement between 4D and KGO. \\
\indent Summing up, we have presented a simple quantum relativistic scheme
to take account the confining potential of quarkonium.  
In this frame we have solved the two-body problem obtaining
the corresponding eigenfuntions and eigenvalues.
We have made a comparison between our model and other relativistic
models. 
This family of mesons entails for a restrictive arena in where
comparison can be made, because 
plenty of models exist with agreement among them and
experimental data. Thus, this can be thought of as a first test of
any alternative setting.
Our results are in good agreement with previous theoretical predictions 
and with experimental data (available in \cite{part}).
This fact allows us to conclude that our treatment reproduces
with a high degree of precision the masses of charmonium and bottonium
family.
In addition, we stress that our model is enterely analytical.\\
\indent We expect that our simple model could be extended to other situations
in the framework of quark models.

\section*{Acknowledgements}
We thank H. Fanchiotti for his constant advice.
We are grateful to C. Na\'on for fruitful discussions. We also
aknoweledge D. Torres for insightful comments and critical reading
of the manuscript.

\begin{table}
\caption{Particular values for $c\bar{c}$ mass spectrum are shown.
In KGO and 4D models, we have considered $m_c=1.5$ GeV and $\omega=0.385$GeV.}

\hfill

\begin{tabular}{cccccccc}     
 & Relativistic Model & \multicolumn{1}{c}{{\em l}=0} 
 & \multicolumn{1}{c}{{\em l}=1}
& \multicolumn{1}{c}{{\em l}=2} & \multicolumn{1}{c}{{\em l}=3} 
& \multicolumn{1}{c}{{\em l}=4} \\ \hline\hline
  &LP &3.105 &3.456 &3.761 &4.034 &4.286 \\
n=0 &QP &3.179 &3.946 &4.193 &4.436 &4.675 \\ 
   &4D &3.531 &3.844 &4.133 &4.404 &4.659 \\
  &{\bf KGO} &3.000 &3.363 &3.691 &3.991 &4.271 \\ \hline
  &LP &3.696 &3.964 &4.215 &4.450 &4.673 \\
n=1 &QP &3.695 &3.946 &4.193 &4.436 &4.675 \\ 
  &4D &4.133 &4.404 &4.659 &4.901 &5.131 \\
  &{\bf KGO} &3.691 &3.991 &4.271 &4.533 &4.781 \\  \hline
  &LP &4.169 &4.397 &4.616 &4.826 & \\
n=2 &QP &4.186 &4.426 &4.662 &4.895 & \\ 
  &4D &4.659 &4.901 &5.131 &5.351 &5.563 \\
  &{\bf KGO} &4.271 &4.533 &4.781 &5.017 &5.242 \\      \hline
  &LP &4.580 &4.783 &4.980 & & \\
n=3 &QP &4.656 &4.886 &5.114 & & \\ 
  &4D &5.131 &5.351 &5.563 &5.767 &5.964 \\
  &{\bf KGO} &4.781 &5.017 &5.242 &5.458 &5.666 
\end{tabular}
\end{table}

\begin{table}
\caption{Particular values for $b\bar{b}$ mass spectrum are shown.
We considered $m_b=5.17$ GeV and $\omega=0.113$GeV.}

\hfill

\begin{tabular}{ccccccccc}        
& Relativistic Model & \multicolumn{1}{c}{{\em l}=0} 
& \multicolumn{1}{c}{{\em l}=1}
& \multicolumn{1}{c}{{\em l}=2} & \multicolumn{1}{c}{{\em l}=3} 
& \multicolumn{1}{c}{{\em l}=4}
\\ \hline\hline
  
n=0 &4D  &10.358 &10.667 &10.795 &10.922 &11.047 \\
  &{\bf KGO} &10.340 &10.452 &10.564 &10.674 &10.782 \\ \hline
n=1 &4D  &10.795 &10.922 &11.047 &11.171 &11.293 \\
  &{\bf KGO} &10.564 &10.674 &10.782 &10.890 &10.997 \\ \hline
n=2 &4D  &10.047 &11.171 &11.293 &11.415 &11.534 \\
  &{\bf KGO} &10.782  &10.890 &10.997 &11.103 &11.208 \\ \hline
n=3 &4D  &11.293 &11.415 &11.534 &11.653 &11.770 \\
  &{\bf KGO} &10.997 &11.103 &11.208 &11.311 &11.414
\end{tabular}
\end{table}

\end{document}